\documentstyle[aps]{revtex}
\input{epsf.tex}
\begin{document}
\draft
\title{Axial collective excitations of a degenerate Fermi gas in the
BEC to unitarity crossover}
\author{R. Combescot and X. Leyronas}
\address{Laboratoire de Physique Statistique,
 Ecole Normale Sup\'erieure*,
24 rue Lhomond, 75231 Paris Cedex 05, France}
\date{Received \today}
\maketitle


\begin{abstract}
We show that, with reasonable hypotheses leading to a simple modeling, a link can be obtained from experiments on the axial low frequency collective modes between the molecular scattering length $a_M$ and the energy parameter $\xi \equiv 1 + \beta$ of the gas at the unitarity limit. We also point out that, in order to reach the range where the features of the Bose limit can be clearly seen, experiments have to go to more dilute situations than have been achieved presently.
\end{abstract}
\vspace{5mm}

Collective excitations of an ultracold dense $^6$Li Fermi gas, most likely in the superfluid regime, have been studied experimentally for the first time very recently \cite{thomas,grim}. In particular Bartenstein \emph{et al} \cite{grim} have covered in part the crossover range going from the unitarity limit to the BEC regime domain \cite{grim}. This is the region to which we will restrict the scope of this paper. In this domain the scattering length $a$ is positive. Physically this is the range where, for a dilute gas, molecules form out of the fermions. At unitarity where $a= \infty$ these molecules are infinitely large and get increasingly small when $a$ decreases. In this last limit these molecules are expected to behave as bosons, and indeed the Bose-Einstein condensation of these molecules has by now been observed by several groups \cite{becmol}. On the other hand this will obviously no longer be true at unitarity. Hence studying the crossover where the internal stucture of the molecules will manifest itself progressively is a very interesting question.

In this paper we consider the detailed information which can be extracted from the experimental results on the low frequency collective modes in this crossover \cite{masa}. We show that, in order to see the specific features of the Bose limit, experiments have to go to more dilute situations than what has been done up to now. With reasonable hypotheses we find that a link can be obtained from experiments between the molecular scattering length $a_M$ and the energy of the gas at the unitarity. Hence experiments could provide $a_M$ together with the first corrective term in the energy to this dilute limit, which is of very high theoretical interest. Moreover we show that, even in the range of dilution which has already essentially been reached, precise experiments can already extract these informations from a detailed analysis of the data. More specifically we will consider in this paper only the quasi one-dimensional situations found for very elongated trap clouds \cite{thomas,grim}. However our approach could be extended to other less anisotropic situations. We restrict also ourselves to the case of atoms being only in two hyperfine states, with equal number of atoms in each hyperfine state. For practical matters we will consider only $^{6}$Li, on which all mode experiments have been performed to date.

In our range of interest we expect superfluid hydrodynamics to be valid because the energy scales set by the molecular binding energy or the pair breaking energy can easily be large, compared to the mode frequencies. In a related way experiments have already reached temperatures low compared to these scales, so that there are few elementary excitations, dissipation can be neglected and we can use perfect fluid hydrodynamics. Indeed dissipation is found to be quite
small \cite{grim} experimentally. Hence this domain is expected to be much simpler than the negative $a$ region, where the pair breaking energy will become smaller as $|a|$ decreases. In particular this will make damping higher, as it is indeed found experimentally. Therefore the validity of reactive
superfluid hydrodynamics is much more uncertain when one goes in the negative $a$ domain.

The mode frequency is determined from the dependence \cite{rmq}  of the chemical potential $ \mu (n) $ on the
total atomic density $n$. More precisely it is related to the derivative of $  \mu (n) $ with respect to $n$ so it conveys detailed and sensitive information on $ \mu (n)$ and on the energy of the gas.  It is conversely possible \cite{rcxl}  to invert
experimental data on the mode frequency to obtain the equation of state of the gas. This
requires only that $ \mu (n)$ is known in some limiting case, from
which one then go away by an iterative procedure, making use of the
experimental knowledge of the mode frequency as a function of
particle number. The basic ingredient of this method has been shown recently
\cite{fuchs} to have an accuracy of order $ 10 ^{-3}$.
However presently the number and the accuracy of the data is not enough to carry out the above program with a sensible precision. Nevertheless we will show that, with reasonable hypotheses, it is possible to analyze the data and extract quite important physical quantities.

We will assume that the (homogeneous) gas can be completely characterized when its density $n$ and the scattering length $a$ are known. This last quantity is controlled experimentally through the applied magnetic field. This hypothesis is far from obvious since one might expect that, at least in the vicinity of the Feshbach resonance, more parameters are needed to fully characterize the scattering properties. However, at least in the case where the width of the resonance is quite large compared \cite{rc} to the Fermi energy, this seems to be valid. This situation is found in $^{6}$Li and in $^{40}$K. Then for homogeneity reasons the $T=0$ chemical potential of an homogeneous gas with density $n$ takes the general form:
\begin{eqnarray}
 \mu(n) = E_{F} f(1/k_F a)
\label{a}
\end{eqnarray}
where $E_{F} = \hbar ^{2} k_F ^{2} /2m $ is the Fermi energy and $ k_F ^{3}=3 \pi ^{2}n $. This implies in particular at resonance $a^{-1}=0$ a universal behaviour since the only energy scale is the Fermi energy. In this case $ \mu(n) =  \xi \, E_{F} $ where $ \xi \equiv 1 + \beta = f(0) $. The energy per atom of the gas itself is given by  $(3/5) \xi \, E_{F} $.

Actually in an harmonic trap the experimental determination of the mode frequency as a function of scattering length allows to obtain $ \mu (n) $ only within a multiplicative constant. This is for example well known for the Bose case where only the linear dependence \cite{str1}  of $ \mu (n) = gn $ on $n$ is necessary to obtain the theoretical result for the frequency, but the constant $ g = 4 \pi \hbar ^{2} a / m$ drops out, so the mode frequency does not allow to reach the scattering length $a$. Similarly in the unitary case \cite{str2} the frequency does not depend on $ \xi$ and one has rather to find $ \xi$ from measurements of the interaction energy of the gas \cite{bourdel}, as it is obtained from expansion experiments. However since there is only a single multiplicative constant, one could hope to find from the mode frequencies the scattering length $a$ as soon as $\xi$ is known, or the reverse. Naturally in order to do so, one needs to make a connection between the regions with small $a$ (in the Bose regime) and with large $a$ (near the unitarity limit). We will make this connection by taking a quite reasonable modeling between these two limiting regions.

In the vicinity of the unitarity limit $ a^{-1} \approx 0$, it is natural \cite{bulg,rcxlcom} to proceed to an expansion of $f(y)$ since $y = 1/k_F a$ is small. Hence we write:
\begin{eqnarray}
f(y) \simeq \xi - S y
\label{b}
\end{eqnarray} 
Near the other limiting case, namely the Bose limit, it is also natural to proceed to an expansion \cite{str2,pit} in powers of the density. The standard expression \cite{str2} of the chemical potential in the Bose limit is  $\mu _{M}= 2 \mu (n) = g_{M}n_{M} $ in terms of the molecular density $n_M$ and the coupling constant is given by $ g_M = 4 \pi \hbar ^{2} a_M / m_M$ in terms of the molecular mass $m_M = 2 m$ and the molecule-molecule scattering length $ a_M$. The next order correction in the standard Bose gas expansion is \cite{pit}  the Lee, Huang and Yang (LHY) term $ (32/3 \sqrt{\pi }) g_{M}n_{M} (n_{M}a_{M}^{3})^{1/2}$. Hence we note that the expansion which comes out naturally in this theoretical framework is in powers of $n^{1/2}$. On the other hand, as we have seen above, it is rather the variable $ k_F a \sim n^{1/3}$ which comes out in studying the BEC-BCS crossover, corresponding physically to the fact that the internal fermionic structure of the Bose molecule under consideration is now coming in. It is unclear whether $n^{1/2}$ terms will be present. At least it is an open theoretical question to find the specific form of the low density expansion, and in particular whether only $n^{1/2}$ or $n^{1/3}$ or both are coming in. In particular experiments on collective modes could contribute to clear up this question. In the present case, since our purpose is to make the connection to the Fermi gas regime, we want naturally to keep the $n^{1/3}$ variable. On the other hand we will not include for simplicity $n^{1/2}$ corrective terms, whose presence is unclear. They are anyway of higher order than $n^{1/3}$. In this way we are led to write the low density expansion of our function $f(y)$ as
\begin{eqnarray}
 f(y) \simeq (A/y) ( 1 + c/y)
 \label{c}
\end{eqnarray}This corresponds to the expansion $ \mu (n) \simeq (1/2) g_{M}n_{M} ( 1 + c (3 \pi ^{2})^{1/3} n^{1/3} a)$, provided we set $ A = (1/3 \pi ) a_M /a $. Naturally parameter $c$ is physically important since it is responsible of the departure of the mode frequency from its ideal Bose gas value in the vicinity of the Bose limit. Similarly parameter $S$ is responsible \cite{rcxlcom} for the linear behaviour \cite{grim} of the mode frequency in the vicinity of unitarity.

We will now connect smoothly the unitarity domain and the Bose domain by making use of a Pad\'e approximation \cite{kimzub} for $f(y)$ with $y$ running in the domain $[0,\infty[$. This kind of approximation is known to be often quite good in reproducing a function because it has the potentiality of imitating the analytical structure of the function by the poles it may introduce. Specifically it amounts to approximate the function of interest by a rational fraction. In the present case it happens that we can make in a natural way such an approximation without introducing more parameters than the four ones we have already discussed. Indeed we will take the following Pad\'e approximation:
\begin{eqnarray}
f(y) = A \, \frac{y+p_0}{y^2 + q_1 y + q_0}
\label{pade}
\end{eqnarray}
Requiring this form to give the proper expansion for small and large $y$ leads to the following expressions for $p_0, q_0$ and $q_1$:
\begin{eqnarray}
p_0 = \xi \, \frac{A+\xi c}{\xi^{2}-AS} \hspace{1cm} q_0= \frac{A p_0}{\xi}  \hspace{1cm} q_1 = p_0 - c
\label{d}
\end{eqnarray}
Naturally behind this approximation we make is hidden the assumption that $ \mu (n)$ is a smooth function of $n$ when one goes from the unitarity to the Bose regime. This means that there is no break in the physical properties during this crossover. Although this is a most common assumption, it is by no means obvious and it is important that it can be justified by detailed experiments on the mode frequencies as a function of scattering length. Indeed it could be then seen if Eq. \ref{pade} gives or not a proper description of these frequencies throughout the whole range of scattering lengths.

We could naturally treat all of our four parameters as free parameters. However this would make our study somewhat heavy. Moreover, in the case of $^{6}$Li, there are already rather reliable informations on the parameters $ \xi $ and $S$. Indeed experimental \cite{bourdel} and theoretical \cite{carl,astr} evaluations of $\xi$ converge around a value $\xi \approx 0.45$. Whatever the final outcome it is not expected to be so far from this value. So we will take $\xi=0.45$ as granted. Similarly the analysis \cite{rcxlcom} of the axial mode frequency data \cite{grim} leads to $S \approx 0.5$, so we will take $ S=0.5$. This leaves us with only two parameters $A$ and $c$.

Parameter $A$ is essentially the molecular scattering length and it is our central interest. It has been shown theoretically \cite{petrov} that  $ a_M = 0.6 a$ in a nearly resonant situation, which corresponds to $A = 0.064$. It would naturally be very interesting to check experimentally this result and we will see that the mode frequency offers a promising way to do it. On the other hand parameter $c$ is the equivalent of the LHY correction in our case and is also clearly of high theoretical interest since it should contain contributions due to the fermionic nature of the molecular components. On theoretical grounds it is unclear to even decide the sign of $c$. However if we take into account the analytical result, to be given below, for the mode frequency near the Bose limit, its extrapolation toward the unitarity region seems to go clearly below the experimental results \cite{grim} if $ c < 0$. We will therefore restrict ourselves to the domain $ c \ge 0$.
\vspace{2cm}
\begin{figure}[htbp]
\begin{center}
\vbox to 50mm { \epsfysize=70mm  \epsfxsize=120mm
\epsfbox{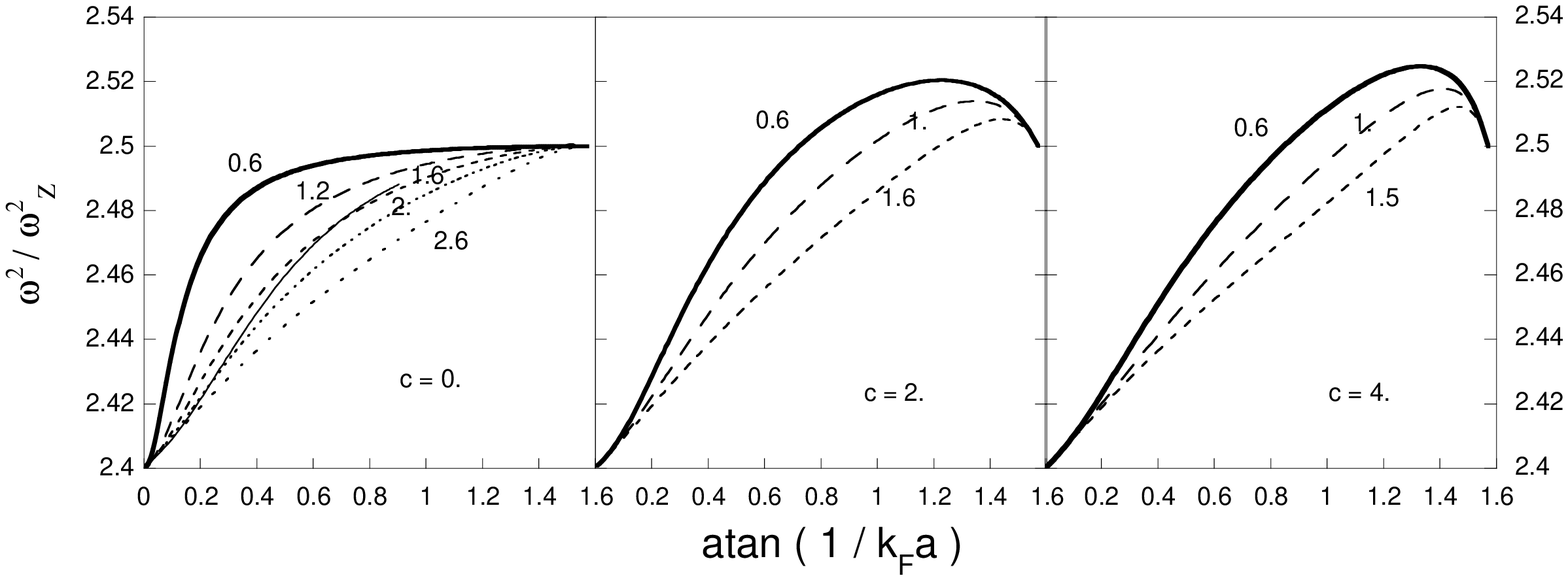} }
\caption{Reduced axial mode frequency squared as a function
of  $\arctan(1/ k_F a )$ for the model Eq. 4 for $c=0,2$ and $4$, and values of $a_M/a$ as indicated near the various curves. For comparison we have also plotted with the $c=0$ case, as the thin line, the result of the model of Ref.13
$f(y)=1/2 - (1/\pi ) \arctan(\pi y/2)$.}
\label{figure1}
\end{center}
\end{figure}
We will now, for consistency, restrict further our parameter domain. First we want $f(y)$ to smoothly interpolate between the unitarity limit where $f(0)= \xi$ and the Bose regime where $f(y) \approx A/y$, so we do not want any zero of $f(y)$ for $ y \ge 0$. This implies $ p_0 > 0$, i.e. $A < 0.4$ or $ a_M/a < 3.8$ (note that in this case we have also $q_0>0$ and $q_1>0$). This relation is well satisfied by all the current possibilities \cite{sdm} for $ a_M/a $ considered in the literature. Next we certainly do not want any pole for $y > 0$. But even poles for $y < 0$ correspond to a nearby singularity which contradicts somewhat our smoothness requirement. Hence we will require that $ f(y) $ has no pole for real $y$ (since we have $q_1 >0$, the real part of the poles location is negative, which puts them away from our $y>0$ domain). This means $ q_{1}^{2}<4q_0$. This condition implies that $A<0.3$ (i.e. $a_M/a < 2.8$) and, in this case, we have the further restriction $ c < (0.73/A) [1+(1-2.47 A)^{3/2}]- 2.7$. 

Before discussing our results, let us first briefly indicate how we proceed to get them and give analytical results in limiting cases. Since we are interested in the axial mode where the gas behaves as a 1D system, we have first \cite{fuchs} to calculate, from the relation between the 3D density $n(\mu )$ and the chemical potential $\mu $, the relation $n_1(\mu )=  \int_{0}^{\mu } d\mu '  n(\mu ')$ between the effective 1D density $n_1$ and the chemical potential $\mu $ on the long axis of  the trap. In our case this calculation can be performed analytically from Eq.(2), but since the result is somewhat lengthy we will not give it explicitely in the general case. On the other hand this relation becomes much simpler near unitarity where $f(y) \simeq  \xi -Sy$. Introducing a parameter $t$ it can be written as $n_1 = t^{5} - (5/8) (S/\xi) t^{4}$, with $\mu = t^{2} -  (S/\xi) t $. In these expressions for $n_1$ and $\mu $, we have actually dropped useless multiplicative factors since they do not enter the final result for the mode frequency. Similarly near the Bose limit where $ f(y) = (A/y) ( 1 + c/y)$, we obtain $n_1 = t^{6} + (8/7) c t^{7}$, with $\mu  = t^{3} +  c t^{4}$. In the general case analogous parametric relations are obtained. The above parameter is actually given by $t= {\bar n}^{1/3}/\kappa $ where $ {\bar n}$ is the (3D) gas density on the long axis of the trap, divided by its maximum value found at the center of the trap, and $\kappa =y_M $ with $y_M$ being the maximal value of $y=1/k_F a$ found at the center of the trap. Hence the maximum value of $t$ is $t_M = k_{Fmax} a$.
\begin{figure}[htbp]
\begin{center}
\hspace{-20mm}
\vbox to 70mm{ \epsfysize=70mm  \epsfxsize=110mm
\epsfbox{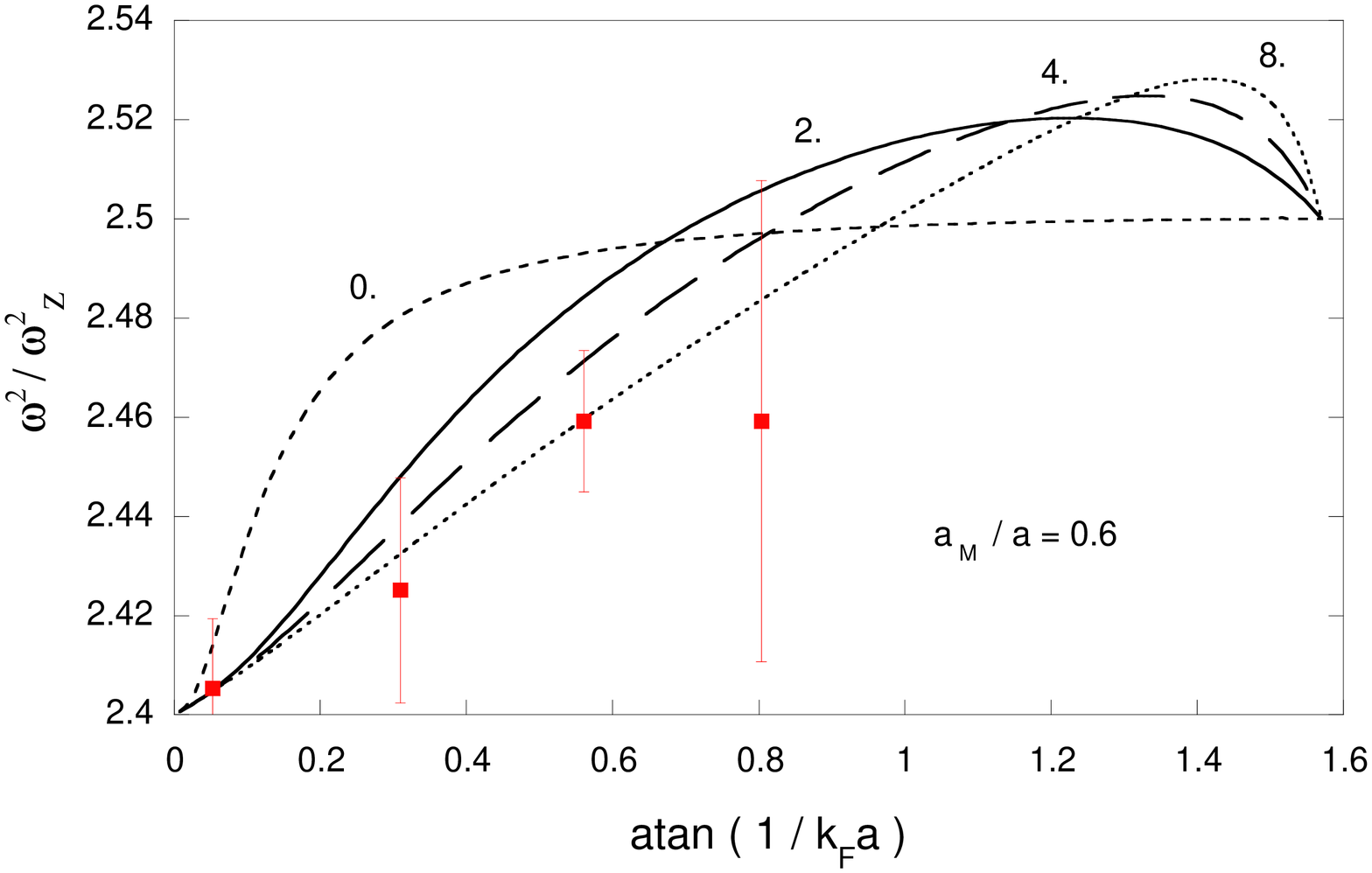} }
\caption{
Reduced axial mode frequency squared as a function
of  $\arctan(1/ k_F a )$ for the model Eq. 4 with the value
 of $a_M/a = 0.6$ of Petrov
\emph{et al} 
and for $c= 0, 2, 4$ and $8$. The filled squares, with error bars, are
the data of Bartenstein \emph{et al}.}
\label{figpetr}
\end{center}
\end{figure}
 We then calculate the mode frequency by making use of the " $\alpha$ -p modeling" together with a first order correction \cite{rcxl,fuchs}. We take the above relation between ${\bar n}_{1} \equiv n_1/n_{1max}$ and ${\bar \mu } \equiv \mu /\mu _{max}$ (with $n_{1max}$ and $\mu _{max}$ being the center trap values of $n_1$ and $\mu $) and make a best approximation of it by $ {\bar n}_1 = [1-(1-{\bar \mu} )^{\alpha /2}]^{p}$. We then make a first order correction to take into account the difference between this model and the actual relation ${\bar n}_{1} ({\bar \mu })$. This procedure leads \cite{fuchs} to a precision of order $10^{-3}$. At unitarity we have the simple result $ {\bar n}_1 = {\bar \mu}^{5/2}$, equivalent to $\alpha =2$ and $p=5/2$, leading \cite{str2}  to $\omega ^{2}/ \omega ^{2}_{z} = 12/5$. Then our first order correction leads, in the vicinity of unitarity, to the analytical \cite{bulg,rcxlcom} result $\omega ^{2}/ \omega^{2}_{z} = 12/5 + (256/875 \pi ) (S/ \xi ) ( 1/ k _{Fmax} a) $. Similarly in the Bose limit we have $ {\bar n}_1 = {\bar \mu}^{2}$, equivalent to $\alpha =2$ and $p=2$. This leads to $\omega ^{2}/ \omega ^{2}_{z} = 5/2$. Then a perturbative calculation gives near this Bose limit $\omega ^{2}/ \omega^{2}_{z} - 5/2 = (3 \sqrt{\pi }/32) (\Gamma (13/3)/ \Gamma (29/6))c (k _{Fmax} a) \simeq 0.082 c (k _{Fmax} a) $. This is to be compared with the coefficient one would obtain \cite{pit} from the LHY result, which can be rewritten as $ f(y) = (A/y) ( 1 + 64 (a_M/6 \pi a)^{3/2}/y^{3/2}) \simeq (A/y) ( 1 + 0.78(a_M/ a)^{3/2}/y^{3/2}) $. The corresponding result \cite{pit}  can be written $\omega ^{2}/ \omega^{2}_{z} - 5/2 = (105/256 \sqrt{6 \pi }) (a_M/ a)^{3/2} (k _{Fmax} a)^{3/2} \simeq  0.094 (a_M/ a)^{3/2} (k _{Fmax} a)^{3/2} $. Hence the numerical coefficient we find in our case, for $c \simeq 1$, is quite similar to the one which comes out in the LHY case.
 
 We give in Fig.\ref{figure1} the results of our numerical calculations for $c=0,2$ and $4$, and values of $a_M/a$ compatible with the restriction given above. In Fig. 2 we take the result $a_M/a = 0.6 $ of Petrov \emph{et al} and give our results for $ c = 0, 2, 4$ and $8$ together with the experimental results. In agreement with our specific choice of $ S/\xi$ all our curves start with the same slope near unitarity. Near the Bose limit our results rise (for $c \neq 0$) above the Bose limit $\omega ^{2}/ \omega^{2}_{z} = 2.5 $. The slope for this rise is naturally proportional to $c$, in agreement with our analytical result. However we notice that this rise is always rather small, in contrast to what one might guess at first. Furthermore we see that values above the Bose limit are essentially only obtained in the density domain beyond the one which has been reached experimentally, so that a positive value for $c$ is not in contradiction with present experimental data. From Fig. 2 we see that $a_M/a = 0.6 $ and $ c \ge 2.$ seems compatible with present experimental data. In this respect we notice that when we translate a dilution parameter \cite{grim}  $n_M a_{M}^{3} = 10^{-3}$, which sounds to correspond very much to a dilute situation, we find (assuming $a_M/a = 0.6 $) $1/k_F a \simeq 1.54$ of order unity, and $ \arctan (1/k_F a) \simeq 1.$ is not in the range where the typical Bose behaviour occurs.
 
 Next we see that, starting from the unitarity value $\omega ^{2}/ \omega^{2}_{z} = 2.4$, the rise toward the Bose value is faster for small $c$ and $a_M/a$. This is easy to understand from Eq. 5 since these small values imply also small values for our parameters $ p_0 , q_0$ and $q_1$. Accordingly the Bose regime, which from Eq. 4 is found for $ y \gg p_0 , q_0, q_1$, is rapidly reached. For larger values of $c$ and $a_M/a$, parameters $ p_0 , q_0$ and $q_1$ get correspondingly larger and the Bose regime is reached only for larger values of $ 1/k_F a$. Nevertheless increasing $c$ or $a_M/a$ does not lead to the same change in the mode frequency, as it can be seen from Fig. 1. Hence by making precise measurements up to $1/k_F a \simeq 2$, it should be possible to obtain both $c$ and $a_M/a$, and even to check if our approximation Eq. 1 of the equation of state gives an accurate description of the data. 
 
In conclusion, by making use of a simple modeling, we have shown how it is possible, from experiments on the axial low frequency collective modes, to find a link  between the molecular scattering length $a_M$ and the energy of the gas at unitarity. We have also pointed out the possibility that the specific features of the Bose limit will be seen in experiments only by going to more dilute situations than what has been done up to now.

We are most grateful C. Salomon for stimulating this work and we acknowledge useful discussions with  F. Chevy, Y. Castin, C. Cohen-Tannoudji and J. Dalibard.

\noindent
* Laboratoire associ\'e au Centre National
de la Recherche Scientifique et aux Universit\'es Paris 6 et Paris 7.



\begin{references}
\bibitem{rmq} Naturally only the density dependent part of the chemical potential enters the mode frequencies. But there are clearly additive contributions to the chemical potential, independent of the density but depending for example on the scattering length. Deep in the Bose regime this will be for example the binding energy $ \hbar^{2}/ma^{2}$ of the molecule. Since such contributions do not affect the mode frequencies, we omit them systematically.
\bibitem{kimzub} Upon writing up this work we have found that J.E. Kim and A.L. Zubarev, cond-mat/0404513 have used also a Pad\'e approximation, but for the energy per particle instead of the chemical potential as we do here.
\end{references}
\end{document}